\documentclass[
  aps,
  prl,
  reprint,
  floatfix,
  superscriptaddress,
  longbibliography
]{revtex4-2}

\usepackage{amsmath,amssymb}
\usepackage{graphicx}
\graphicspath{{figures/}}

\begin{document}

\title{Superselectivity as a Receptor-Fluctuation Response}

\author{Xiuyang Xia}
\email{xiaxy@sustech.edu.cn}
\affiliation{Department of Physics, Southern University of Science and Technology, Shenzhen, Guangdong 518055, China}
\affiliation{Center for Complex Flows and Soft Matter Research, Southern University of Science and Technology, Shenzhen, Guangdong 518055, China}

\date{}

\begin{abstract}
Superselective multivalent binding enables sharp receptor-density discrimination in targeting and sensing, but what its logarithmic response measures microscopically remains unclear.
Here, we derive an exact response decomposition of the selectivity into receptor-state reweighting and a direct occupation response.
In the dilute quenched Poisson limit, the direct term vanishes and selectivity is exactly the mean receptor excess beneath bound particles, which can be measured at a single density from a sufficiently large co-registered receptor--particle image.
Beyond Poisson statistics, the excess is normalized by the Fano factor for a linear count response, while changes in distribution shape require the full count-resolved response.
Lattice simulations verify these relations across distinct receptor statistics, while off-lattice simulations show that particle crowding enters through the direct term.
These results establish a fluctuation--response framework linking multivalent selectivity to the local receptor fluctuations preferentially sampled by adsorption.
\end{abstract}

\maketitle

Multivalent particles distribute binding across many weak and reversible contacts \cite{Mammen1998,Fasting2012,Huskens2006}, and the number of bonding arrangements can grow rapidly as more receptors become accessible \cite{Kitov2003,Varilly2012,AngiolettiUberti2013}.
This combinatorial amplification, termed superselectivity, was proposed as a design principle that allows adsorption to switch on within a narrow receptor-density range while each microscopic bond remains reversible \cite{MVF2011}.
Subsequent work demonstrated superselectivity in multivalent polymers and supramolecular assemblies \cite{Dubacheva2014,Dubacheva2015,Albertazzi2013} and visualized it directly in DNA-mediated colloids and influenza-virus adhesion \cite{Linne2021,Overeem2021}.
Experiments and theory conventionally quantify superselectivity through the logarithmic response of adsorbed coverage to the mean receptor density \cite{DubachevaReview2023}.

However, a bound particle samples one local receptor configuration, whereas the logarithmic response depends on how a density scan reweights the full ensemble of configurations.
Depending on the receptor statistics, increasing the mean density may shift the count distribution, broaden its tails, alter the weight of pre-existing dense patches, or reorganize correlated receptors, so comparable enrichment beneath adsorbed particles can be associated with different selectivities \cite{Liu2020,Dubacheva2019,XiaUniformity2023,Xie2025}.
A microscopic theory must therefore connect the receptor configurations selected by adsorption to the way those configurations are reweighted along the density scan.

The same distinction limits current measurements because selectivity is usually reconstructed from a receptor-density titration by fitting or smoothing the adsorption curve and then taking a local derivative \cite{Linne2024,Ahnert2007}.
The resulting uncertainty is especially acute at the low particle activities used to suppress adsorption saturation and expose superselectivity, where few bound particles are available for measurement.
Even under these conditions, co-registered receptor--particle images preserve the local receptor environments associated with binding, information that is discarded when the data are reduced to an adsorption curve \cite{Linne2021,Overeem2021}.

Here, we derive an exact response decomposition that separates the receptor contribution to selectivity from changes in occupation at a fixed receptor state.
For dilute adsorption on a quenched landscape, the latter vanishes, and the Poisson limit reduces selectivity to the mean receptor excess beneath bound particles.
We test the framework across receptor ensembles with distinct fluctuation statistics and in continuous-space simulations, linking macroscopic selectivity to spatially resolved receptor and particle statistics.

\paragraph*{Theory.---}
We consider multivalent particles adsorbing onto a surface with a spatially fluctuating receptor landscape, where each particle samples the receptors within a binding footprint of fixed area $A_0$.
The mean receptor count per footprint is $\bar n=\rho_{\rm R}A_0$, where $\rho_{\rm R}$ is the mean receptor density, and adsorption is quantified by the dimensionless coverage $\Theta=A_0\langle N_{\rm P}\rangle/A$, with $A$ the surface area and $N_{\rm P}$ the number of adsorbed particles.
We assume that the local occupation probability depends on the receptor state only through its count $n$ and write the coverage as $\Theta(\bar n)=\sum_{n=0}^{\infty}P(n;\bar n)\theta(n;\bar n)$, where $P(n;\bar n)$ is the count distribution and $\theta(n;\bar n)$ is the conditional occupation probability.
This expression also defines the receptor ensemble observed beneath adsorbed particles because random footprints sample $P(n;\bar n)$, whereas footprints centered on adsorbed particles sample $P_{\rm occ}(n;\bar n)=P(n;\bar n)\theta(n;\bar n)/\Theta(\bar n)$, making the shift from $P$ to $P_{\rm occ}$ the microscopic sampling bias imposed by adsorption.

We vary $\bar n$ while holding fixed the dimensionless particle activity per footprint $z$, ligand valency $\kappa$, single-bond statistical weight $s$, footprint area $A_0$, and receptor-ensemble parameters.
The selectivity is the logarithmic response $\alpha\equiv(\partial\ln\Theta/\partial\ln\bar n)_{\mathcal C}$, where $\mathcal C$ denotes these fixed conditions.
Along this scan, the count-resolved response $\mathcal R(n;\bar n)\equiv\partial\ln P(n;\bar n)/\partial\ln\bar n$ is the state-resolved observable conjugate to $\ln\bar n$, quantifying how the statistical weight of each receptor count changes.
Because $\Theta$ normalizes the occupied weight $P(n;\bar n)\theta(n;\bar n)$, its logarithmic derivative is the occupied-footprint average of the logarithmic response of this weight, and separating the receptor and occupation factors gives
\begin{equation}
\alpha=\left\langle\mathcal R(n;\bar n)\right\rangle_{\rm occ}+\left\langle\frac{\partial\ln\theta(n;\bar n)}{\partial\ln\bar n}\right\rangle_{\rm occ},
\label{eq:full_response_decomposition}
\end{equation}
where $\langle\cdot\rangle_{\rm occ}$ averages over receptor counts in occupied footprints.
Eq.~\ref{eq:full_response_decomposition} resolves the macroscopic derivative into the reweighting of receptor states selected by adsorption and the direct change in occupation at a fixed local receptor state (see Supplemental Material~\cite{SupplementalMaterial} for the complete derivation).
For quenched receptors and dilute particles, changing $\bar n$ reweights receptor states without altering their conditional occupation probabilities at fixed $n$.
The direct occupation term therefore vanishes, leaving $\alpha=\langle\mathcal R(n;\bar n)\rangle_{\rm occ}$.
Since normalization of $P(n;\bar n)$ gives $\langle\mathcal R(n;\bar n)\rangle=0$ for random footprints, the receptor contribution measures the excess density response of the receptor environments selected by adsorption.

\paragraph*{Poisson receptor fluctuations.---}
\begin{figure}[t]
\includegraphics[width=\columnwidth]{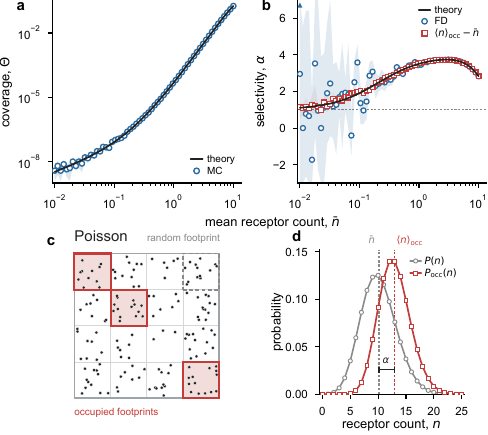}
\caption{Poisson receptor fluctuations and selectivity.
Panels (a,b) compare coverage and selectivity from lattice-gas Monte Carlo (MC) simulations with exact theory.
Panel (b) shows selectivity from finite differences (FD) and from the occupied-footprint receptor excess $\langle n\rangle_{\rm occ}-\bar n$.
Panels (c,d) show a Poisson receptor map and the corresponding random- and occupied-footprint count distributions at $\bar n=10.14$.
The separation between their mean counts equals $\alpha$.
The adsorption parameters are $\kappa=4$, $s=\mathrm e^2$, and $z=10^{-8}$.
Shading denotes 95\% intervals from complete-replica bootstrap resampling; boundary triangles mark zero lower bounds or confidence sets extending beyond the plotted range.}
\label{fig:poisson}
\end{figure}
We first consider the common Poisson limit, which applies when excluded-volume effects and receptor--receptor interactions are negligible on the scale of a binding footprint, as on randomly grafted model surfaces or in fixed receptor snapshots with negligible correlations on the scale of $A_0$ \cite{RogersCrocker2011,WangDNA2015}.
The receptor number $n$ within a footprint then follows Poisson statistics, for which the count-resolved response is $\mathcal R(n;\bar n)=n-\bar n$.
Substituting this response into the quenched relation gives
\begin{equation}
\alpha=\langle n\rangle_{\rm occ}-\bar n.
\label{eq:poisson_local_excess}
\end{equation}
Increasing $\bar n$ reweights a footprint containing $n$ receptors at the logarithmic rate $n-\bar n$, and averaging this rate over particle-centered footprints makes the macroscopic response equal to their mean receptor excess relative to random footprints (see Supplemental Material~\cite{SupplementalMaterial}).
In the Poisson limit, superselectivity therefore means that each occupied footprint selects, on average, more than one excess receptor.

Eq.~\ref{eq:poisson_local_excess} gives a single-condition estimator from a sufficiently large co-registered image because random windows of area $A_0$ determine $\bar n$, while identical particle-centered windows determine $\langle n\rangle_{\rm occ}$.
Their difference gives $\alpha$ without differentiating across density conditions, although near adsorption onset the field of view must contain enough occupied footprints to control the sampling uncertainty.

\begin{figure*}[t]
\includegraphics[width=\textwidth]{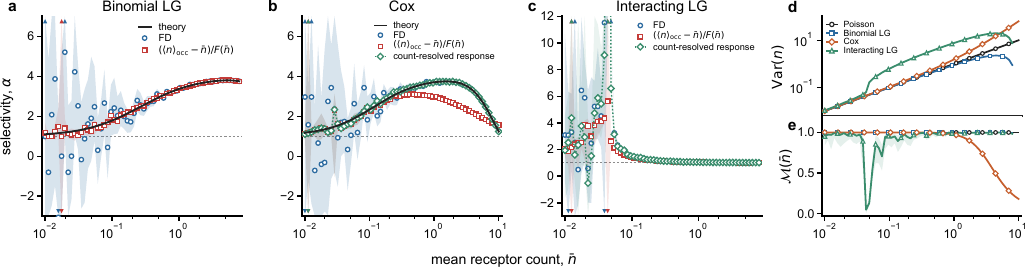}
\caption{Receptor statistics and selectivity.
Panels (a--c) compare selectivity from finite differences (FD), the Fano-rescaled receptor excess, and the count-resolved response for finite-capacity binomial, frozen-intensity Cox, and attractive interacting receptor ensembles.
Panels (d,e) show the corresponding receptor-count variance and linear count-mode weight $\mathcal M(\bar n)$.
The binomial ensemble has a footprint capacity $M=9$.
In the binary Cox ensemble, the local intensity multiplier $X$ equals $0.4$ with probability $0.6$ and $1.9$ otherwise.
The interacting lattice gas uses a dimensionless nearest-neighbor attraction $J=3$.
The adsorption parameters are $\kappa=4$, $s=\mathrm e^2$, and $z=10^{-8}$.
Shaded regions denote 95\% complete-replica bootstrap intervals, and boundary triangles indicate intervals extending beyond the plotted range.}
\label{fig:receptor_statistics}
\end{figure*}

We tested Eq.~\ref{eq:poisson_local_excess} in an independent-site multivalent lattice-gas model on a $400\times400$ surface divided into binding zones, each assigned a quenched Poisson receptor count and allowed to host at most one tetravalent particle.
The Monte Carlo coverage in Fig.~\ref{fig:poisson}(a) agrees with the exact Poisson-averaged lattice-gas result.
As shown in Fig.~\ref{fig:poisson}(b), the finite-difference selectivity follows the exact result at appreciable coverage but becomes noisy near adsorption onset, whereas the local receptor excess follows the exact curve throughout the scan.
This comparison confirms that the receptor environments selected by adsorbed particles determine the density response directly from local statistics.

\paragraph*{Beyond Poisson.---}
Beyond Poisson statistics, the same receptor excess can produce a different selectivity because the background count fluctuations have a different amplitude.
We quantify this amplitude by the footprint Fano factor $F(\bar n)\equiv\langle(n-\bar n)^2\rangle/\bar n$.
Increasing $\bar n$ reweights the footprint ensemble toward larger counts on balance, so the correlation between this reweighting and the count excess must reproduce the imposed change in the mean count, $\langle(n-\bar n)\mathcal R(n;\bar n)\rangle=\partial\langle n\rangle/\partial\ln\bar n=\bar n$.
When this reweighting is linear in $n$, the identity gives $\mathcal R(n;\bar n)=(n-\bar n)/F(\bar n)$, and averaging over occupied footprints yields
\begin{equation}
\alpha=
\frac{\langle n\rangle_{\rm occ}-\bar n}{F(\bar n)},
\label{eq:fano_rescaled_excess}
\end{equation}
which converts the receptor excess selected by adsorption into a density response using the fluctuation amplitude of the background landscape (see Supplemental Material~\cite{SupplementalMaterial} for the linear-mode derivation).

Eq.~\ref{eq:fano_rescaled_excess} applies when increasing $\bar n$ tilts the count distribution toward larger $n$ in proportion to the count excess, whereas heterogeneity or collective receptor organization can reweight the tails or distinct receptor populations nonlinearly.
We distinguish these responses by the fraction of the density sensitivity captured by the count excess,
\begin{equation}
\mathcal M(\bar n)\equiv
\frac{\bar n}{F(\bar n)\langle\mathcal R(n;\bar n)^2\rangle},
\label{eq:linear_count_mode_weight}
\end{equation}
in which the numerator is the sensitivity required to shift the mean count, while $\langle\mathcal R(n;\bar n)^2\rangle$ is the full Fisher information of the count distribution with respect to $\ln\bar n$, so $\mathcal M(\bar n)$ measures how much of the overall statistical change is visible through receptor enrichment alone.
All averages are over random footprints, and the mean-count identity gives $\mathcal M(\bar n)=\operatorname{Corr}^{2}[n,\mathcal R(n;\bar n)]$ and hence $0\leq\mathcal M(\bar n)\leq1$ (see Supplemental Material~\cite{SupplementalMaterial}).
The limit $\mathcal M(\bar n)=1$ means that the receptor contribution to selectivity is fully captured by the Fano-rescaled count excess, whereas $\mathcal M(\bar n)<1$ reveals nonlinear redistribution of receptor-count probabilities that must be retained through the full $\mathcal R(n;\bar n)$.
The distribution-level origin of these linear and nonlinear responses is shown in Fig.~S1 of the Supplemental Material~\cite{SupplementalMaterial}.
Together, Eqs.~\ref{eq:full_response_decomposition} and \ref{eq:linear_count_mode_weight} separate the origin of selectivity at two levels: receptor-state reweighting from direct occupation response, and within the receptor contribution, count enrichment from nonlinear changes in distribution shape.
Experimentally, receptor-count histograms across a receptor-only density series determine $\mathcal M(\bar n)$ and, when needed, $\mathcal R(n;\bar n)$, while a co-registered image at the target density supplies the occupied-footprint average.

Finite-capacity binomial statistics provide the simplest non-Poisson case with a linear count response.
When site-occupancy correlations are weak, this model represents a molecular scaffold or confined membrane region with a limited number of binding-competent sites \cite{ShawNanocalipers2014,Veneziano2020,Salaita2010}, with receptor uniformity providing an experimentally controllable parameter for multivalent adsorption \cite{XiaUniformity2023}.
Each footprint contains $M$ independently occupiable receptor sites, giving the Fano factor $F(\bar n)=1-\bar n/M$ and a linear response for which Eq.~\ref{eq:fano_rescaled_excess} remains exact despite the sub-Poissonian fluctuations.
The Fano-rescaled receptor excess in Fig.~\ref{fig:receptor_statistics}(a) follows the exact binomial result and agrees with the finite-difference estimate within uncertainty, while the finite-difference points scatter near the low-coverage onset.
The suppressed variance in Fig.~\ref{fig:receptor_statistics}(d) changes the response scale, and $\mathcal M(\bar n)=1$ in Fig.~\ref{fig:receptor_statistics}(e) confirms that the linear count mode carries the full response.

Frozen local-intensity heterogeneity provides a distinct route to nonlinear count response, which we model with a Cox ensemble whose receptor count in each footprint is Poisson distributed with a quenched local mean $\bar nX$ \cite{Cox1955,Moller1998}.
Such frozen variations in receptor availability can arise from membrane organization or cytoskeletal confinement \cite{Eggeling2009,DouglassVale2005,Lillemeier2010} and from steric modulation by the glycocalyx \cite{Paszek2014,Stone2017}.
Because a high receptor count reflects both the local background intensity and the global receptor density, $\mathcal R(n;\bar n)$ becomes nonlinear in $n$, and the Fano-rescaled receptor excess consequently departs from the exact Cox response at high density in Fig.~\ref{fig:receptor_statistics}(b).
The occupied-footprint average of the full $\mathcal R(n;\bar n)$ follows the exact result and remains consistent with the finite-difference selectivity within uncertainty, while the simultaneous decrease of $\mathcal M(\bar n)$ below unity in Fig.~\ref{fig:receptor_statistics}(e) identifies the missing nonlinear response modes.

Attractive receptor interactions provide another route to nonlinear count response, which we model with a square-lattice gas at a dimensionless nearest-neighbor attraction $J=3$, well above the critical value \cite{LeeYang1952}, as a coarse-grained representation of receptor-poor and receptor-rich membrane regions generated by clustering or phase separation \cite{BanjadeRosen2014,Su2016,Dubacheva2019,Xie2025}.
Changing $\bar n$ at fixed $J$ transfers statistical weight between these regions and reshapes the footprint-count distribution.
Near phase separation, the Fano-rescaled receptor excess in Fig.~\ref{fig:receptor_statistics}(c) captures only the variance rescaling and departs from the finite-difference selectivity, whereas the full count-resolved response detects the nonlinear transfer of weight.
Because the finite-size phase conversion lies within one sampled density interval, both discrete derivative reconstructions are sensitive to density resolution at the two conversion points, although the count-resolved response follows the finite-difference selectivity within uncertainty away from this narrow interval.
The enhanced variance in Fig.~\ref{fig:receptor_statistics}(d) measures the strength of the count fluctuations, while the drop of $\mathcal M(\bar n)$ in Fig.~\ref{fig:receptor_statistics}(e) identifies their nonlinear reorganization.
Full density scans and equilibration checks for the interacting receptor landscapes are provided in Fig.~S2 of the Supplemental Material~\cite{SupplementalMaterial}.

\paragraph*{Beyond independent adsorption.---}
\begin{figure}[t]
\includegraphics[width=\columnwidth]{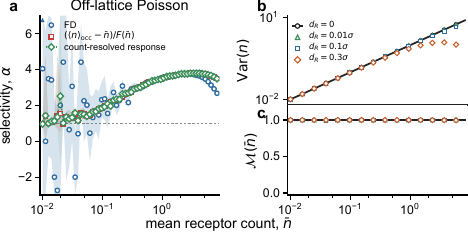}
\caption{Receptor exclusion and particle crowding.
Panel (a) compares point-receptor selectivity from finite differences (FD), the Fano-rescaled receptor excess, and the count-resolved response.
The receptor-based measurements overlap but separate from FD at high density because of guest-particle crowding.
Panels (b,c) show the footprint-count variance and linear count-mode weight $\mathcal M(\bar n)$ for $d_{\rm R}/\sigma=0$, $0.01$, $0.1$, and $0.3$.
Here $d_{\rm R}$ is the receptor hard-core diameter and $\sigma$ is the guest-particle diameter.
The adsorption parameters are $\kappa=4$, $s=\mathrm e^2$, and $z=10^{-8}$.
Shaded regions denote 95\% complete-replica bootstrap intervals, and boundary triangles indicate intervals extending beyond the plotted range.}
\label{fig:off_lattice}
\end{figure}
Finally, we examine how receptor exclusion and particle crowding modify the fluctuation response using multivalent hard disks of diameter $\sigma$ adsorbing on two-dimensional receptor fields in continuous space.
At each receptor density, we sample a receptor configuration and then fix the receptor positions during particle adsorption.
Point receptors recover the off-lattice Poisson limit, while finite receptor diameter $d_{\rm R}$ introduces steric anticorrelations and increasing particle coverage eventually couples neighboring binding footprints through excluded volume.
These effects enter Eq.~\ref{eq:full_response_decomposition} through different physical channels, as receptor exclusion changes receptor-state reweighting whereas particle crowding generates a direct occupation response at fixed local receptor count.

For point receptors, the Poisson count distribution gives $\mathcal M(\bar n)=1$, so the Fano-rescaled excess and full count-resolved response overlap in Fig.~\ref{fig:off_lattice}(a), as confirmed by Fig.~\ref{fig:off_lattice}(c).
At low particle coverage, both receptor-based measurements agree within uncertainty with the finite-difference selectivity, confirming the Poisson/Fano relation in continuous space using random circular windows for the background statistics and identical particle-centered windows for the occupied statistics.
At high $\bar n$, the finite-difference selectivity bends below the two receptor-based measurements because neighboring hard disks restrict the available adsorption configurations, making the occupation probability at a fixed receptor count density dependent and producing a negative direct occupation response.

Finite receptor size produces a separate receptor effect, as increasing $d_{\rm R}$ suppresses the footprint-count variance in Fig.~\ref{fig:off_lattice}(b) while $\mathcal M(\bar n)$ remains close to unity in Fig.~\ref{fig:off_lattice}(c).
Receptor exclusion therefore acts mainly by rescaling the linear count fluctuation, while larger $d_{\rm R}$ also lowers the guest coverage reached at fixed $\bar n$ and weakens the high-density crowding contribution.
The finite-diameter response curves and their coverage-dependent separation are shown in Fig.~S3 of the Supplemental Material~\cite{SupplementalMaterial}.

\paragraph*{Discussion.---}
Our results identify the receptor fluctuations selected by bound particles as the microscopic origin of logarithmic selectivity in dilute multivalent adsorption.
At low particle coverage on fixed receptor landscapes, selectivity is the occupied-footprint average of the receptor density response, giving microscopic content to the conventional combinatorial picture and implying that superselectivity on a Poisson landscape corresponds to more than one excess receptor per occupied footprint.
In this regime, superselectivity measures the density response of the receptor fluctuations selected by multivalent binding.
The full decomposition extends this receptor-centered relation by retaining changes in occupation at a fixed receptor state, which the off-lattice simulations trace to guest-particle crowding at high coverage.

The framework suggests an experimental diagnostic in which spatially resolved measurements compare background receptor fluctuations with those sampled by bound particles in targeted adhesion and biosensing platforms \cite{Carlson2007,AngiolettiUberti2017,Squires2008}.
In practice, receptor statistics are set by how receptors are presented and organized \cite{DubachevaReview2023}, with the Poisson excess applying to random grafting at low surface density, Fano rescaling applying when finite site capacity or modest correlations mainly alter the fluctuation amplitude, and the full count-resolved response becoming necessary when clustering or spatial heterogeneity reshapes the distribution.
Receptor organization therefore controls not only the binding curve but also the fluctuation modes through which multivalent recognition becomes superselective.
The same separation may guide extensions to multicomponent receptor fields \cite{Curk2017}, nonequilibrium adsorption \cite{Ravnik2026}, deformable interfaces \cite{Amjad2017}, and image-based inference \cite{Jungmann2016,Levet2019}.

\begin{acknowledgments}
This work was supported by the Fundamental and Interdisciplinary Disciplines Breakthrough Plan of the Ministry of Education of China (Project No.~JYB2025XDXM505).
\end{acknowledgments}

\bibliography{manuscript}

\end{document}

\title{Supplemental Material for ``Superselectivity as a Receptor-Fluctuation Response''}

\author{Xiuyang Xia}
\email{xiaxy@sustech.edu.cn}
\affiliation{Department of Physics, Southern University of Science and Technology, Shenzhen, Guangdong 518055, China}
\affiliation{Center for Complex Flows and Soft Matter Research, Southern University of Science and Technology, Shenzhen, Guangdong 518055, China}

\date{}

\maketitle

\section{General theory for quenched receptor landscapes}
\label{sec:general_theory}

\subsection{Local binding law and quenched average}
\label{sec:local_binding}

We consider a particle that samples a binding footprint of fixed area \(A_0\), within which the mean receptor count is
\begin{equation}
\bar n=\rho_{\rm R}A_0.
\label{eq:mean_count}
\end{equation}
Because the receptor landscape remains fixed during the binding measurement, a random footprint samples the probability mass function \(P(n;\bar n)\), where \(n=0,1,2,\ldots\) and \(\langle n\rangle=\bar n\).

For the lattice models used in the main text, a particle has a nominal ligand valency \(\kappa\), and a state with \(b\) ligand--receptor bonds has the combinatorial weight \(\binom{\kappa}{b}(n)_b\), where the falling factorial \((n)_b=n!/(n-b)!\) vanishes when \(b>n\).
With a single-bond statistical weight \(s\), the local bound-state partition sum is
\begin{equation}
q(n)=\sum_{b=1}^{\min(\kappa,n)}\binom{\kappa}{b}(n)_b s^b.
\label{eq:local_partition_sum}
\end{equation}
If \(z\) is the dimensionless particle activity associated with one binding footprint, the corresponding single-footprint occupation law is
\begin{equation}
\theta(n)=\frac{zq(n)}{1+zq(n)}.
\label{eq:local_occupation}
\end{equation}
Eq.~\ref{eq:local_occupation} applies when guest particles occupy independent footprints, while the more general notation \(\theta(n;\bar n)\) allows the guest-particle environment at fixed \(n\) to change along the receptor-density scan.

The adsorption coverage is the quenched average of the local occupation law,
\begin{equation}
\Theta(\bar n)=\sum_{n=0}^{\infty}P(n;\bar n)\theta(n;\bar n).
\label{eq:quenched_coverage}
\end{equation}
In Eq.~\ref{eq:quenched_coverage}, the local binding problem is evaluated at fixed \(n\) before averaging over receptor fluctuations, and the same weighting defines the receptor-count distribution sampled by occupied footprints as
\begin{equation}
P_{\rm occ}(n;\bar n)=\frac{P(n;\bar n)\theta(n;\bar n)}{\Theta(\bar n)}.
\label{eq:occupied_distribution}
\end{equation}
All occupied-footprint averages below use the conditional distribution in Eq.~\ref{eq:occupied_distribution}.

\subsection{Count-resolved response identity}
\label{sec:response_identity}

The count-resolved response function measures how the probability of a particular receptor count changes with the mean count,
\begin{equation}
\mathcal R(n;\bar n)=\frac{\partial\ln P(n;\bar n)}{\partial\ln\bar n}.
\label{eq:response_definition}
\end{equation}
For a receptor ensemble parametrized by \(\ln\bar n\), \(\mathcal R(n;\bar n)\) is the state-resolved observable conjugate to this parameter.
When the parameter enters an equilibrium Boltzmann weight as a thermodynamic field, the same logarithmic derivative reduces to the centered conjugate variable familiar from conventional fluctuation-response relations; the present definition remains valid for an arbitrary receptor-count family.
Taking the density derivative at fixed $z$, $\kappa$, $s$, $A_0$, and receptor-ensemble parameters and differentiating Eq.~\ref{eq:quenched_coverage} without assuming independent guest particles gives
\begin{equation}
\alpha=\left\langle\mathcal R(n;\bar n)\right\rangle_{\rm occ}+\left\langle\frac{\partial\ln\theta(n;\bar n)}{\partial\ln\bar n}\right\rangle_{\rm occ}.
\label{eq:explicit_local_response}
\end{equation}
Eq.~\ref{eq:explicit_local_response} resolves the selectivity into the reweighting of receptor-count fluctuations and the direct change in the local occupation probability at fixed \(n\).
When guest particles occupy approximately independent footprints and the local binding parameters remain fixed, the conditional occupation probability at a fixed receptor state is unchanged along the density scan.
The direct term then vanishes, and the response reduces to
\begin{equation}
\alpha=\left\langle\mathcal R(n;\bar n)\right\rangle_{\rm occ}.
\label{eq:central_identity}
\end{equation}
Eq.~\ref{eq:central_identity} is exact in the independent-guest limit for any receptor-count distribution and any fixed local occupation law.
Normalization of \(P(n;\bar n)\) gives the corresponding random-footprint average
\begin{equation}
\left\langle\mathcal R(n;\bar n)\right\rangle=0.
\label{eq:response_zero_mean}
\end{equation}
Eq.~\ref{eq:central_identity} is therefore the excess density response of the receptor states selected by adsorption relative to those sampled by random footprints.

Reducing the local receptor state to \(n\) assumes that this count contains all local information relevant to binding; if receptor geometry affects adsorption at fixed \(n\), Eq.~\ref{eq:central_identity} remains valid after replacing the count by the complete local receptor state, although a count histogram alone no longer contains the required response.

\subsection{Linear count mode and nonlinear response}
\label{sec:linear_mode}

Differentiating the mean-count constraint with respect to \(\ln\bar n\) gives
\begin{equation}
\left\langle(n-\bar n)\mathcal R(n;\bar n)\right\rangle=\bar n.
\label{eq:response_mean_constraint}
\end{equation}
The footprint Fano factor is defined by
\begin{equation}
F(\bar n)=\frac{\left\langle(n-\bar n)^2\right\rangle}{\bar n}.
\label{eq:footprint_fano}
\end{equation}
Eqs.~\ref{eq:response_mean_constraint} and \ref{eq:footprint_fano} determine the projection of the count-resolved response onto the linear fluctuation \(n-\bar n\).
The remaining response is orthogonal to that fluctuation because
\begin{equation}
\left\langle(n-\bar n)\left[\mathcal R(n;\bar n)-\frac{n-\bar n}{F(\bar n)}\right]\right\rangle=0.
\label{eq:response_orthogonality}
\end{equation}
Averaging the linear projection and the orthogonal remainder over occupied footprints gives the decomposition used in the main text,
\begin{equation}
\alpha=\frac{\langle n\rangle_{\rm occ}-\bar n}{F(\bar n)}+\left\langle\mathcal R(n;\bar n)-\frac{n-\bar n}{F(\bar n)}\right\rangle_{\rm occ}.
\label{eq:fano_decomposition}
\end{equation}
Eq.~\ref{eq:fano_decomposition} separates the Fano-rescaled receptor excess from density-dependent changes in distribution shape that cannot be represented by a linear count fluctuation.

We quantify the fraction of the Fisher information carried by the linear projection as
\begin{equation}
\mathcal M(\bar n)=\frac{\bar n}{F(\bar n)\left\langle\mathcal R(n;\bar n)^2\right\rangle}.
\label{eq:linear_mode_weight}
\end{equation}
Because \(\langle\mathcal R(n;\bar n)\rangle=0\), Eq.~\ref{eq:linear_mode_weight} can equivalently be written as \(\mathcal M(\bar n)=\operatorname{Corr}^{2}[n,\mathcal R(n;\bar n)]\).
Because \(\langle\mathcal R(n;\bar n)^2\rangle\) is the Fisher information of the receptor-count distribution with respect to \(\ln\bar n\), the Cauchy--Schwarz inequality and Eq.~\ref{eq:response_mean_constraint} imply \(0\leq\mathcal M(\bar n)\leq1\).
The limit \(\mathcal M(\bar n)=1\) means that the count-resolved response is linear on the support of \(P(n;\bar n)\), so the nonlinear term in Eq.~\ref{eq:fano_decomposition} vanishes for every local binding law.
A value below unity identifies additional response modes but does not represent the fraction of a particular selectivity curve explained by receptor counts.

\subsection{Reconstruction from receptor and particle images}
\label{sec:image_reconstruction}

Once a receptor-only calibration establishes that the response is linear, a sufficiently large co-registered receptor--particle image gives a single-condition estimator in which random windows determine \(\bar n\) and \(F(\bar n)\), while particle-centered windows of the same size determine \(\langle n\rangle_{\rm occ}\).
The Fano-rescaled difference between these means gives the selectivity, provided that the field of view contains enough occupied footprints to estimate the particle-centered mean with the desired precision.

For a general receptor-count family, the response can be reconstructed from random-window histograms at neighboring mean densities.
Taking \(\bar n\) as the geometric midpoint of \(\bar n_-\) and \(\bar n_+\), count bins with nonzero support at both neighboring densities give the symmetric log-ratio estimator
\begin{equation}
\widehat{\mathcal R}(n;\bar n)=\frac{\ln\widehat P(n;\bar n_+)-\ln\widehat P(n;\bar n_-)}{\ln(\bar n_+/\bar n_-)}.
\label{eq:log_ratio_response}
\end{equation}
The leading spacing error in Eq.~\ref{eq:log_ratio_response} is quadratic in the separation in \(\ln\bar n\).
The reconstructed selectivity is obtained by averaging the response values over the occupied count histogram,
\begin{equation}
\widehat\alpha(\bar n)=\sum_n\widehat P_{\rm occ}(n;\bar n)\widehat{\mathcal R}(n;\bar n).
\label{eq:histogram_selectivity}
\end{equation}
This procedure differentiates receptor-count probabilities sampled by many local windows rather than a sparse adsorption curve.

For the interacting lattice-gas calculations, we used the equivalent centered difference of the probabilities themselves,
\begin{equation}
\widehat{\mathcal R}(n;\bar n)=\frac{\widehat P(n;\bar n_+)-\widehat P(n;\bar n_-)}{\widehat P(n;\bar n)\ln(\bar n_+/\bar n_-)}.
\label{eq:probability_difference_response}
\end{equation}
Eq.~\ref{eq:probability_difference_response} avoids taking the logarithm of an empty neighboring bin and converges to Eq.~\ref{eq:response_definition} as the density spacing vanishes.
Bins with zero probability at the central density carry no weight in Eq.~\ref{eq:histogram_selectivity}.

In experimental images, rare-count bins may be pooled or regularized before taking a logarithm, while the retained count range must have support in both neighboring receptor histograms and non-negligible occupied probability at the central density.
Because overlapping windows are spatially correlated, uncertainty should be estimated by resampling complete images or spatial blocks, and a receptor-density gradient or adjacent patterned regions can provide \(\bar n_-\), \(\bar n\), and \(\bar n_+\) under closely matched imaging and surface-chemistry conditions.

\subsection{Locality and particle crowding}
\label{sec:locality}

Eq.~\ref{eq:central_identity} assumes that the occupation probability of a footprint is determined by its local receptor state and remains fixed along the density scan, a condition that can fail even for an exactly known Poisson landscape when adsorbed particles exclude one another.
At appreciable particle coverage, the occupation of a candidate footprint depends on neighboring particles whose distribution also changes with \(\bar n\), producing a direct particle-ensemble response that is absent from \(\mathcal R(n;\bar n)\) and cannot be recovered by measuring the receptor-count distribution more accurately.
The local estimator therefore applies when particle crowding and longer-ranged particle correlations are negligible, while outside this regime the local state must be enlarged to include the relevant particle environment.

\section{Response for specific receptor-count distributions}
\label{sec:distributions}
\begin{figure}[p]
\centering
\includegraphics[width=\textwidth]{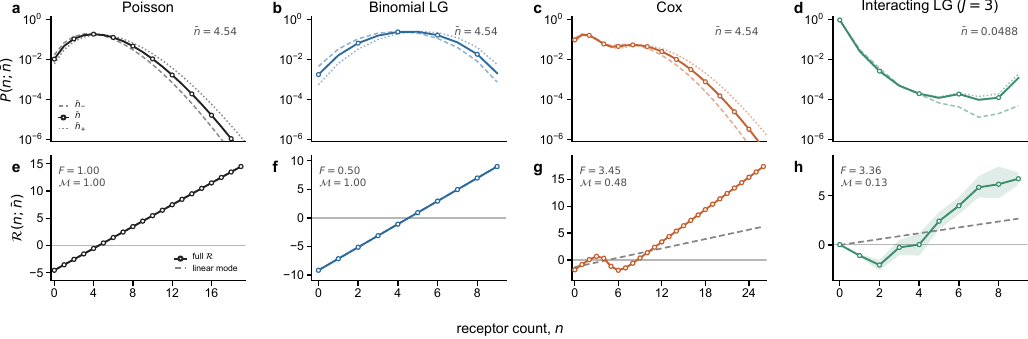}
\caption{Anatomy of the receptor-count response.
Panels (a--d) show neighboring count distributions for Poisson, finite-capacity binomial, binary Cox, and interacting lattice-gas receptor ensembles, while panels (e--h) show the corresponding full count-resolved responses at the central density.
The gray dashed line denotes the linear count mode \((n-\bar n)/F(\bar n)\).
The full response coincides with this mode for the Poisson and binomial ensembles, whereas Cox heterogeneity and interacting receptors generate nonlinear components.
The first three columns are exact at \(\bar n=4.54\), while panels (d,h) use the \(J=3\) production histograms at \(\bar n=0.0488\), with the shaded region in panel (h) denoting the 95\% bootstrap interval over ten receptor replicas.}
\label{fig:response_anatomy}
\end{figure}

Figure~\ref{fig:response_anatomy} compares how the four receptor ensembles are reweighted along a density scan, and the following subsections derive the response represented in each column.

\subsection{Poisson receptor counts}
\label{sec:poisson_distribution}

Independent receptors give the Poisson receptor-count distribution
\begin{equation}
P(n;\bar n)=\ee^{-\bar n}\frac{\bar n^n}{n!}.
\label{eq:poisson_distribution}
\end{equation}
Differentiating Eq.~\ref{eq:poisson_distribution} gives the linear count-resolved response
\begin{equation}
\mathcal R(n;\bar n)=n-\bar n.
\label{eq:poisson_response}
\end{equation}
The footprint Fano factor is
\begin{equation}
F(\bar n)=1.
\label{eq:poisson_fano}
\end{equation}
The central identity therefore reduces to
\begin{equation}
\alpha=\langle n\rangle_{\rm occ}-\bar n.
\label{eq:poisson_selectivity}
\end{equation}
Eq.~\ref{eq:poisson_selectivity} remains valid at finite particle activity because local occupation saturation is already contained in the occupied distribution.
Consequently, no additional factor involving \(1-\Theta\) is required, and as every local occupation probability approaches unity, the occupied distribution approaches the random-footprint distribution so that both the receptor excess and selectivity vanish.

\subsection{Finite-capacity binomial receptor counts}
\label{sec:binomial_distribution}

Suppose a footprint contains \(M\) independently occupiable receptor sites and that each site is occupied with probability \(\bar n/M\).
The receptor-count distribution is
\begin{equation}
P(n;\bar n)=\binom{M}{n}\left(\frac{\bar n}{M}\right)^n\left(1-\frac{\bar n}{M}\right)^{M-n}.
\label{eq:binomial_distribution}
\end{equation}
The finite capacity suppresses the count variance and gives
\begin{equation}
F(\bar n)=1-\frac{\bar n}{M}.
\label{eq:binomial_fano}
\end{equation}
The corresponding response remains linear,
\begin{equation}
\mathcal R(n;\bar n)=\frac{n-\bar n}{1-\bar n/M}.
\label{eq:binomial_response}
\end{equation}
The selectivity is therefore
\begin{equation}
\alpha=\frac{\langle n\rangle_{\rm occ}-\bar n}{1-\bar n/M}.
\label{eq:binomial_selectivity}
\end{equation}
The binomial result, evaluated with \(M=9\) in the main text, shows that non-Poisson variance alone does not require a nonlinear correction.

\subsection{Binary Cox receptor counts}
\label{sec:cox_distribution}

A Cox process represents quenched variations of the local Poisson intensity, which we write as \(\bar nX\) using a density-independent multiplier \(X\) with unit mean \cite{Cox1955,Moller1998}.
Because the footprint count is Poisson conditional on \(X\), its marginal distribution is
\begin{equation}
P(n;\bar n)=\sum_a p_a\ee^{-\bar n x_a}\frac{(\bar n x_a)^n}{n!}.
\label{eq:cox_distribution}
\end{equation}
Here \(X=x_a\) with probability \(p_a\), and the count-resolved response depends on the posterior mean intensity at fixed count according to
\begin{equation}
\mathcal R(n;\bar n)=n-\bar n\langle X\mid n\rangle.
\label{eq:cox_response}
\end{equation}
The footprint Fano factor follows from the law of total variance,
\begin{equation}
F(\bar n)=1+\bar n\Var(X).
\label{eq:cox_fano}
\end{equation}
The selectivity can be written in terms of occupied-footprint averages as
\begin{equation}
\alpha=\langle n\rangle_{\rm occ}-\bar n\langle X\rangle_{\rm occ}.
\label{eq:cox_selectivity}
\end{equation}
Eq.~\ref{eq:cox_selectivity} separates receptor enrichment from the density response because adsorption favors footprints with a high pre-existing intensity \(X\), whereas changing \(\bar n\) rescales the intensity within both receptor-poor and receptor-rich environments.

The binary Cox ensemble in the main text uses \(p_{\rm low}=0.6\), \(x_{\rm low}=0.4\), and \(x_{\rm high}=1.9\), which give \(\langle X\rangle=1\) and \(\Var(X)=0.54\).
The posterior \(\langle X\mid n\rangle\) varies nonlinearly with \(n\), so \(\mathcal M(\bar n)<1\) and the Fano-rescaled receptor excess is not exact.
Because the model specifies only one-footprint count statistics, it has no spatial patch size or correlation length and does not represent domain formation or phase separation.

\subsection{Interacting receptors and phase conversion}
\label{sec:interacting_distribution}

The interacting receptor model is a canonical square-lattice gas with nearest-neighbor attraction \(J\), whose dimensionless receptor energy for site occupations \(r_i\) is
\begin{equation}
\mathcal H_{\rm R}=-J\sum_{\langle ij\rangle}r_i r_j.
\label{eq:receptor_lattice_energy}
\end{equation}
Kawasaki exchange conserves the total receptor number and samples the equilibrium receptor ensemble at each prescribed \(\bar n\), while periodically translated \(3\times3\) windows define footprint counts with a maximum \(M=9\).
The calculations compare the exact square-lattice critical coupling \(J_c=2\ln(1+\sqrt{2})\) with the strong-attraction value \(J=3\) used in the main text.

At strong attraction, the receptor-count distribution near phase separation can be viewed approximately as a mixture of receptor-poor and receptor-rich window distributions,
\begin{equation}
P(n;\bar n)=[1-\phi(\bar n)]P_{\rm poor}(n)+\phi(\bar n)P_{\rm rich}(n).
\label{eq:two_phase_distribution}
\end{equation}
Here \(\phi(\bar n)\) is the area fraction of the receptor-rich phase, and if the two within-phase distributions change slowly across the coexistence region, transfer of statistical weight between them dominates the count-resolved response,
\begin{equation}
\mathcal R(n;\bar n)=\frac{\partial\phi}{\partial\ln\bar n}\frac{P_{\rm rich}(n)-P_{\rm poor}(n)}{P(n;\bar n)}.
\label{eq:phase_conversion_response}
\end{equation}
The ratio in Eq.~\ref{eq:phase_conversion_response} is generally nonlinear in \(n\), with interface windows, finite within-phase widths, and finite-size changes in the two component distributions adding further deviations from a linear response.

The \(J=3\) calculations in the main text use the measured count histograms rather than the approximate two-phase form in Eq.~\ref{eq:two_phase_distribution}, with the full response reconstructed from neighboring densities using Eq.~\ref{eq:probability_difference_response}.
The large count variance measures the collective receptor fluctuations, while the reduction of \(\mathcal M(\bar n)\) identifies redistribution among receptor-poor, receptor-rich, and interfacial count states.
The underlying identity concerns a derivative at fixed \(\bar n\), whereas both the adsorption derivative and the histogram response are reconstructed from a finite density interval that contains the sharp conversion between receptor-poor and receptor-rich configurations in the finite \(J=3\) system.
The two reconstructed responses are therefore resolution sensitive at the conversion points and should not be interpreted as pointwise estimates of the continuum derivative there.

\subsection{Off-lattice hard-disk receptor fields}
\label{sec:hard_disk_distributions}

The off-lattice calculations use circular particle footprints of area \(A_0=\pi\sigma^2/4\), where \(\sigma\) is the adsorbed-particle diameter, and point receptors form a Poisson process whose random-footprint counts obey Eq.~\ref{eq:poisson_distribution}.
Giving the receptors a hard-core diameter \(d_{\rm R}\) introduces anticorrelations that suppress the footprint variance, so the resulting count distribution is sampled directly from random circular windows and its response is reconstructed from neighboring receptor densities.

For the diameters \(d_{\rm R}=0.01\sigma\), \(0.1\sigma\), and \(0.3\sigma\) used in the main text, \(\mathcal M(\bar n)\) remains close to unity, decreasing to \(0.93\) only at the most densely packed \(d_{\rm R}=0.3\sigma\) state, where the full count response separates weakly from the Fano approximation.
The growing separation between the local receptor response and finite-difference selectivity at increasing particle coverage instead identifies guest-particle crowding as the dominant high-density correction.
Because increasing \(d_{\rm R}\) lowers the guest coverage reached at fixed \(\bar n\), this high-density separation is smaller for the largest receptors.

\section{Simulation protocols and uncertainty estimates}
\label{sec:simulation_uncertainty}

\subsection{Quenched lattice simulations}
\label{sec:quenched_numerics}

The independent Poisson, binomial, and binary Cox calculations use a \(400\times400\) array of binding zones, each containing a quenched receptor count and hosting at most one particle.
The local model uses \(\kappa=4\), \(s=\ee^2\), and the dimensionless footprint activity \(z=10^{-8}\), with 10 independent receptor landscapes and 1,000 independent particle-occupation states sampled per landscape at each \(\bar n\).
The displayed scan contains 60 logarithmically spaced densities together with one additional density on either side for centered finite differences.

The interacting calculations use a \(400\times400\) receptor lattice with periodic boundaries and a fixed total receptor number.
Every receptor configuration contributes all periodically translated \(3\times3\) footprint counts.
The main \(J=3\) scan uses 10 independently equilibrated receptor configurations and 1,000 particle-occupation samples per configuration, while the check at \(\bar n=4.5\) uses 20 receptor replicas.
Global occupied--vacant exchange moves preserve the receptor number and sample the canonical lattice gas, while the \(J=3\) production configurations use coexistence-informed droplet, slab, or bubble initial states followed by 32,000 or 64,000 burn-in sweeps depending on density.
A separate 128,000-sweep preflight monitors the bond density, low-wavevector amplitude, largest-cluster fraction, and \(3\times3\)-window Fano factor, whereas the random-start stress test at half filling remains limited by slow domain coarsening and is not pooled with production data.

\subsection{Off-lattice hard-disk simulations}
\label{sec:off_lattice_numerics}

The off-lattice simulations contain adsorbed hard-disk particles of diameter \(\sigma\) and quenched fields of point receptors or hard-disk receptors with \(d_{\rm R}/\sigma=0.01\), \(0.1\), and \(0.3\).
The particle binding radius is \(\sigma/2\), giving the footprint area \(A_0=\pi\sigma^2/4\), and the model includes receptor--receptor and particle--particle hard-core exclusion but no particle--receptor steric exclusion.
Receptors are prepared independently at each density and held fixed during particle sampling, with each finite-diameter field equilibrated for 1,000 canonical displacement sweeps before random windows and particle configurations are sampled.
Particle insertion uses the local binding weights in Eq.~\ref{eq:local_partition_sum} with \(\kappa=4\), \(s=\ee^2\), and an areal activity \(z_{\rm 2D}\) chosen so that \(z_{\rm 2D}A_0=z=10^{-8}\), matching the dimensionless footprint activity of the lattice model.
The scan contains 60 logarithmically spaced values from \(\bar n=0.01\) to \(8\) together with one neighboring density on either side for centered finite differences, and each density uses 10 complete replicas whose 40 periodic tiles span 160,000 footprint areas.
For each tile, particle grand-canonical Monte Carlo consists of \(5\times10^5\) equilibration moves followed by \(2\times10^7\) production moves, with one configuration recorded every \(2\times10^4\) moves and the 40 tiles aggregated to give 1,000 particle samples per complete replica.
Random circular windows determine \(P(n;\bar n)\), \(\mathrm{Var}(n)\), and \(\mathcal R(n;\bar n)\), while particle-centered windows determine the occupied count distribution.

\begin{figure}[p]
\centering
\includegraphics[width=\textwidth]{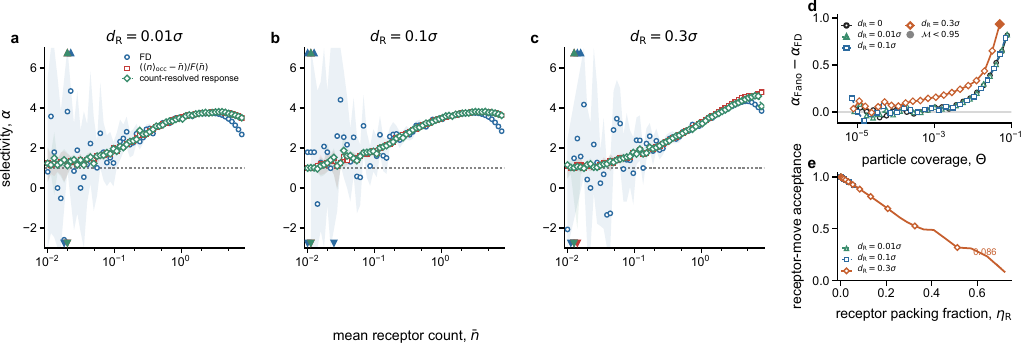}
\caption{Finite receptor size and particle crowding in the off-lattice model.
Panels (a--c) compare selectivity from finite differences, the Fano-rescaled excess, and the full count response for finite receptor diameters, with shading denoting 95\% complete-replica bootstrap intervals and boundary triangles denoting intervals extending beyond the displayed range.
Panel (d) shows the difference between the local Fano response and finite-difference selectivity versus particle coverage for states with at least \(10^4\) occupied observations, with the filled symbol marking \(\mathcal M<0.95\).
Panel (e) shows receptor-move acceptance during preparation of the quenched fields.}
\label{fig:off_lattice_crowding}
\end{figure}

The preparation diagnostic in Fig.~\ref{fig:off_lattice_crowding}(e) remains finite throughout the scan but falls to \(0.086\) at receptor packing fraction \(0.72\), making this final state the most demanding quenched field in the data set.

\subsection{Finite differences and uncertainty estimates}
\label{sec:uncertainty}

Finite-difference selectivities use the two neighboring density states without fitting or smoothing the adsorption curve, and uncertainty is evaluated using 50,000 bootstrap draws that resample only complete receptor landscapes.
Tiles, overlapping windows, and particle configurations within a complete replica are not treated as independent bootstrap units, while replicas with no adsorbed particles and bootstrap draws with no occupation information are retained without introducing a pseudocount.

Coverage and receptor-count variance use 95\% percentile intervals of complete-replica bootstrap means.
Selectivity intervals are obtained by basic-bootstrap inversion of denominator-free estimating equations rather than by directly resampling a logarithm or ratio, with the finite-difference, Fano-rescaled, and count-resolved responses recomputed within each draw.
The linear count-mode weight is likewise reconstructed from resampled neighboring count distributions rather than from a fixed fitted response.

Shaded regions in the main and Supplemental figures denote these 95\% intervals, while a triangle at a plot boundary indicates that the confidence set extends beyond the displayed range rather than ending at a clipped finite value.
For an uninformative bootstrap draw, the allowed response set is retained rather than deleting the draw.

\bibliography{manuscript}

\clearpage

\clearpage
\begin{figure}[p]
\centering
\includegraphics[width=\textwidth]{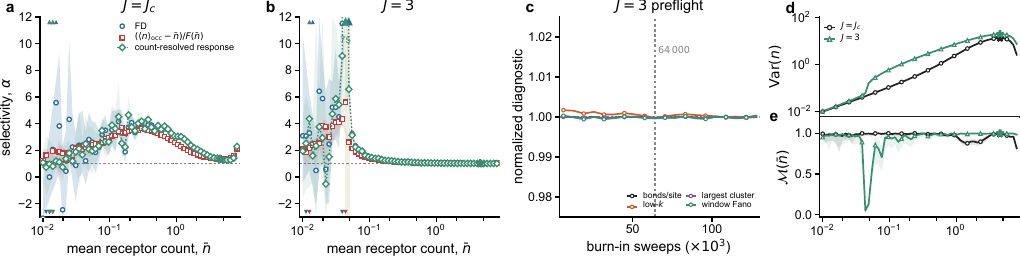}
\caption{Interacting-receptor scans and numerical validation.
Panels (a,b) compare selectivity from finite differences, the Fano-rescaled excess, and the count-resolved response across the complete \(J=J_c\) and \(J=3\) density scans, with shading denoting 95\% complete-replica bootstrap intervals.
Boundary triangles denote intervals extending beyond the displayed range, while stars mark independent 20-replica checks at \(\bar n=4.5\).
The vertical band in panel (b) contains the finite-size phase conversion, and the count-resolved point above the plotting range is labeled by its central value.
Panel (c) shows equilibration diagnostics for the \(J=3\), \(\bar n=4.5\) preflight normalized by their terminal-quarter means, with bands showing standard errors over five receptor replicas and the vertical line marking 64,000 burn-in sweeps.
Panels (d,e) show the footprint-count variance and linear count-mode weight for the two interaction strengths.}
\label{fig:interacting_validation}
\end{figure}

\clearpage
\clearpage